\documentclass[12pt,preprint]{aastex}
\def\lax {\ifmmode{_<\atop^{\sim}}\else{${_<\atop^{\sim}}$}\fi}
\def\gax {\ifmmode{_>\atop^{\sim}}\else{${_>\atop^{\sim}}$}\fi}
\def\gtorder{\mathrel{\raise.3ex\hbox{$>$}\mkern-14mu
    \lower0.6ex\hbox{$\sim$}}} 

\begin{document}

\title{New Evidence for a Black Hole in the Compact Binary
  Cygnus X-3}

\author{Chris R. Shrader,\altaffilmark{1, 2},   Lev
  Titarchuk\altaffilmark{1, 3} and Nikolai Shaposhnikov \altaffilmark{1,4} }

\altaffiltext{1}{Goddard Space Flight Center, NASA,  Astrophysics
  Science Division, code 661, Greenbelt MD 20771, 
  Chris.R.Shrader@nasa.gov}

\altaffiltext{2}{Universities Space Research Association, 10211
  Wincopin Cir, Suite 500, Columbia MD, 21044 }

\altaffiltext{3}{University of Ferrara, Italy, Goddard Space Flight
  Center, NASA,  code 663, Greenbelt   MD 20771;
  George Mason University/Center for Earth
  Observing and Space Research, Fairfax, VA 22030;  US Naval Research
  Laboratory, Code 7655, Washington, DC 20375-5352;
  lev.titarchuk@nrl.navy.mil}

\altaffiltext{4}{Department of Astronomy, University of Maryland, College
 Park MD 20742}  

\begin{abstract}
The bright and highly variable X-ray and radio source known as Cygnus
X-3 was among the first X-ray sources discovered, yet it remains in
many ways an enigma. Its known to consist of a massive, Wolf-Rayet
primary in an extremely tight orbit with a compact object. Yet one of
the most basic  of parameters - the mass of the compact object - is
not known. Nor is it  even clear whether its is a neutron star or a
black hole. In this Paper we present our analysis of the broad-band
high-energy continua covering a substantial  range in luminosity and
spectral morphology. We apply these results to a recently identified
scaling relationship which has been demonstrated to provide reliable
estimates of the  compact object mass in a number of accretion powered
binaries.  This analysis leads us to conclude that the compact 
object in Cygnus X-3 has a mass greater than $4.2M_\odot$
thus clearly indicative of a black hole and as such
resolving a long-standing issue. The full range of uncertainty in our 
analysis and from using a range of recently published distance estimates
 constrains the compact object mass to lie between
$4.2M_\odot$ and $14.4M_\odot$. Our favored estimate, based on a
9.0 kpc distance estimate is $\sim 10 M_\odot$ with the 
error margin of 3.2 solar masses. 
This result may thus pose challenges to shared-envelope
evolutionary models of compact binaries, as well as establishing Cygnus
X-3 as the first confirmed accretion-powered galactic gamma-ray source.

\end{abstract}


\section{Introduction}
The bright X-ray binary source known as Cygnus X-3 (herein Cyg X-3)
was discovered over 40 years ago  by \cite{gia67}, and has been
the subsequent focus of extensive study. It nonetheless remains
enigmatic in that the nature of the compact object; a neutron star or
black hole has not been unambiguously determined, its high-amplitude
intensity and spectral variations in the X-ray and radio bands are not
well understood, and its low-amplitude and featureless power-density
spectrum does not resemble other known neutron star or black hole
X-ray binaries, e.g. \cite{McC99}. Many of these unique 
attributes are due to the nature of the donor star,
widely believed to be a Wolf Rayet (WR) star, e.g. \cite{vankerk92} . 
It is also the strongest radio source among X-ray binaries,
with a quiescent flux $<100$ mJy, a flux of 0.1 Jy flaring to several
Jy in high-activity states. It has recently been detected in the
~0.1-1-GeV domain with the Fermi Gamma-Ray Space Telescope \citep{abdo09}
and is thus among a small number of gamma-ray bright X-ray binaries.  
Its  orbital period of 4.8 hours is typical for a low-mass
binary, but its WR donor star could have a mass upwards of 30
M$_\odot$, thus making it an extremely tight binary system, all the 
more so if the compact star is a black hole.  

Distance estimates to Cyg X-3 range from 7.2 to  9.3 kpc;
see details in \cite{lzt09}. Here we adopt the distance of 9 kpc, 
inferred from dust-halo scattering measurements \citep{pred00}
however, we consider the full range of these distance determinations
in evaluating the uncertainty of our result. 
Such a large distance means that it is viewed
thorough many magnitudes of visual extinction. It is in addition
likely to be enshrouded in the dense wind environment of the WR donor
star. This combination of high local and Galactic line-of-sight column 
densities has rendered the usual photometric and spectroscopic
techniques for deriving a binary solution intractable, and as a
result, no reliable mass estimate for the compact has emerged. 
The local absorption has also limited conclusions based on X-ray
spectroscopy, which may be further complicated by 
 the presence of a synchrotron component associated with the radio 
and gamma-ray emission and the effects of Compton downscattering.  
While hard-to-soft spectral transitions suggestive of a black hole
binary are evident, e.g. \cite{hz09}
they cannot always be easily 
reconciled within the context of
familiar black-hole low-hard to high-soft state transition patterns, nor do
they resemble neutron-star Z or atoll source spectral behavior. No
pulsations are detected, which would be a clear signature of a
magnetized neutron star, however, it is possible that pulsed (radio or
X-ray) emission could be present in the system but unobservable due to
the effects of dense ambient plasma. It is noted that this type of
scenario has been proposed and is considered credible in explanation
of another radio-loud, gamma-ray-bright X-ray binary LSI +61 303. 

Previous  estimates of the compact object mass based on a variety of methods 
have been attempted,  but the results tend to be highly uncertain.  For 
example Schmutz, Geballe and Schild (1996) employing IR spectroscopy obtained 
a likely mass in the range of $7-40 M_\odot$ with a best estimate of 
$17 M_\odot$. Hanson, 
Still and Fender (2000) constructed a radial velocity curve from which derive 
a mass function of $0.027 M_\odot$, which leads to a blackhole mass of less than 
$10 M_\odot$ but could still accommodate a neutron star depending on the true
value of the binary inclination. Vilhu et al. (2009) assess 
orbital modulation of X-ray emission lines and find a most likely compact object 
masses between 2 and $8 M_\odot$ while Hjalmarsditter et al. (2008) estimate a 
$\sim 30 M_\odot$ black hole based on interpretation of color-luminosity diagrams. 

Our approach involves a multi-epoch spectral analysis and assessment
of the inferred parameters.
\cite{st09} and \cite{ts09}, hereafter  ST09 and TS09 respectively
established well defined correlations between certain temporal and 
spectral parameters. Furthermore, these 
correlation curves observed in different sources 
exhibited differing patterns for different binaries, but that
the variety of those patterns was limited, and thus scalable.
Recently, the application of this method was extended to study of 
an another class of X-ray source, the
ULX NGC 5408 X-1 (Strohmayer and Mushotzky 2009). That study 
resulted in tangible evidence for the existence of an intermediate mass 
object in that system.

The ubiquitous nature of  these correlations led to 
the suggestion that the underlying
physical processes leading to the observed variability properties
are closely tied to  the Comptonizing media. Furthermore, they vary in a well
defined  manner as the source makes a transition between spectral
states.  The fact that the same correlations are seen in many sources,
which vary widely in both luminosity (and thus presumably with mass accretion
rate) and spectral state, suggests a common set of underlying physical 
conditions.
TS09 showed that in GRS 1915+105 the photon index monotonically
increases with  disk mass accretion rate, followed by saturation.
The radio 
luminosity does not correlate with low frequency quasi-periodic 
oscillations or X-ray luminosity for the
entire range of spectral states, from low-hard to high-soft through
intermediate.

In this Paper we present the results of a broad-band
multi-epoch spectral analysis of Cyg X-3, applied to a novel approach for
compact object mass determination, using photon index$-$disk mass
accretion rate correlation first implemented by  ST09.   We compiled a
data base of some 35 observations of Cyg X-3, primarily from the RXTE
satellite, applied similar modeling methods, and studied
the inferred parameter interdependencies. From the results of that
analysis, we have applied the scaling laws derived in ST09 deriving an
estimate of the compact object mass, which we find to be consistent
with a black hole.

In \S 2 we briefly discuss our methodology for mass-determination,
which is described in detail elsewhere. In \S 3 we describe our
database compilation and analysis techniques, in \S 4 we present  our
results, draw comparison to similar results previously calibrated from
objects with independent mass estimates, and in \S 5 we summarize our
conclusions.

\section{Data Analysis and Modeling}
\subsection{Data Selection}

We extracted data from the HEASARC archives covering an approximately
9-year span starting at modified Julian day 50319 (where MJD=JD-2,400,000.5). 
RXTE was our
primary database, but we also examined several epochs of INTEGRAL
and BeppoSAX coverage as well, but ultimately chose to use
the RXTE data almost exclusively in our final mass-estimation
analysis. Some 35 spectra were examined, 
spanning a large range of hardness and intensity variations. Figure 1
depicts the distribution of these data in hardness-intensity space. As
is well known, Cyg X-3 undergoes extreme spectral evolution the full range
of which cannot be represented by a single model, e.g. \cite{hz09}. As
we describe in detail in the next section, we found satisfactory
fits to our model for the subset of the data which were not at the 
extreme range of Figure 1, i.e the very hard and very soft states of the
system. 

The data reduction and analysis was performed using current HEADAS
software release (version 6.7) and the RXTE instrumental response file
release in 2009, August. For our spectral analysis we use the PCA
Standard-2 data mode and the Standard Archive HEXTE Mode. Standard
dead time correction was applied to all spectra. Spectra were modeled
using the XSPEC analysis package. We used approximate energy
intervals of 3.0-30.0 keV and 20.0-200.0 keV  for the PCA and HEXTE data
correspondingly, although depending on the spectral and intensity
state in question additional channel selection cuts were made as
warranted. A systematic error of 1 percent was assumed
to account for uncertainty of the absolute instrumental calibration.
 The stated uncertainties on spectral model parameters were
calculated using 1-$\sigma$ confidence interval. 

\subsection{Model Fitting}

We examined each individual observation applying our spectral
fitting model which we describe below. We note that in many instances,
we were simply unable to achieve a satisfactory fit to the data. The
complex variations of the local absorption are likely the underlying
issue here, as has been pointed out by others; see e.g. \cite{zdz09}.
This is consistent at least with Figure 1, where the subset of the data
we ultimately selected for our mass-determination analysis is depicted.
The hardness-intensity space is roughly divided there into extreme-soft,
moderate, and extreme hard spectral regimes. In the latter, the effects
of Comptonization from the ambient plasma or from electron populations 
associated with a collimated outflow or with coronal electrons lead
to spectral hardening. For the extreme soft case the hard powerlaw
is very poorly constrained (or undetected). An example is depicted in 
Figure 2, which is based on an INTEGRAL observation. We note that although 
the soft-X-ray detector on INTEGRAL, JEM-X, is much less sensitive than the RXTE
PCA, the IBIS hard X-ray detector is superior in the hard X-ray regime.
Clearly the hard-powerlaw is not reliably determined. Furthermore, with 
the bolometric luminosity is strongly skewed to the ~keV regime, uncertainties
in the flux and spectral shape due to absorption are substantial.

We do however obtain satisfactory fits to our model for a
subset of about half 
of the observations studied. In these cases the reduced
$\chi^2$ statistics are of order unity resulted and meaningful constraints
on the key parameters of our model were achieved. These data are 
represented in the hardness-intensity space by the blue points in Figure 1,
which span a range of about 0.1-0.35 in hardness defined by 
the ratio of PCA counts in the $(9-20)~{\rm keV}/(2-9)~{\rm keV}$ bands. 
The rationale for this selection is as follows. The extreme soft-state 
spectra cannot be used for index-mass accretion
rate correlation. As noted, this is because these emergent 
spectra are a result of reprocessing of
the emitted spectra in the optically thick cloud
covering the central source. In this case, the entirety of information
on the nature of the central source is lost. In the extreme hard-state
case, we speculate that an additional Comptonization component associated
with plasma ambient to the central accretion flow contributes to
a hardening and curvature of the continuum above a 10 keV. This is
counter to the basic converging inflow scenario we propose.
The specific
data sets included in our mass-determination analysis, along with the
the inferred spectral parameters are enumerated in Table 1.

Our spectral model, consists of so called, the bulk-motion Comptonization (BMC) model;
see  \cite{bmc}, hereafter TMK97, convolved with model 
corrections for (Galactic plus local)
absorption, Fe K$_\alpha$ emission and absorption edge, and an
exponential cutoff. The BMC model is a convolution an input thermal
spectrum having normalization $N_{bmc}$ and color temperature $kT_c$
with a Green's function encoding the Comptonization
process which is a broken power law. This approach is applicable 
to any type of Comptonization process
The generic shape of a Comptonization Green function is a 
broken power law which is followed by exponential cutoff at energies higher 
than the kinetic energy of matter and it 
is independent of the type of Comptonization process, e.g. bulk or 
thermal. An example of our model fitted to 
the data is depicted in Figure 3.

As an aside note, this Green's function technique is 
well suited to the analysis of
a wide variety of problems, and has been applied for example to 
a description of the income distribution in financial analysis by \cite{tls09}.    
The normalization $N_{bmc}$ is a ratio of the source luminosity to the
square of the distance. The spectral index parameter of the
model has been found to correlate approximately linearly with the
normalization up to a saturation level, typically about 2.2$-$2.8,
after which it flattens.  For various sources the same effect is
seen in photon index $\Gamma$ vs. quasi-periodic oscillation frequency
(see ST09).
Furthermore, the self-similar nature of an ensemble of observations
can be related by scaling laws. 

The  resulting spectra are characterized by the inferred parameters
$\log(A)$ related to the Comptonized fraction $f$ as $f=A/(1+A)$ and to a
spectral index $\alpha=\Gamma-1$.  There are  several advantages of
using the BMC model with respect to the more common approach, i.e. 
the application of an additive combination of a
blackbody/multi-color-disk plus power-law/thermal
Comptonization. First, by its nature, it is
applicable to the general case of photon energy gain through
not only thermal Comptonization but also via dynamic (bulk) motion
Comptonization \citep[see][for details]{ST06}. Second, the BMC spectral 
shape  has an appropriate low energy curvature, which is not the case with
a simple additive power law. This is essential for a correct
representation of the lower energy spectrum. This can lead for example to
inaccurate $N_H$ column values 
and thus produce an non-physical component ``conspiracy'' 
involving the high-energy spectral cutoff and spectral-index 
parameterizations.
Specifically, when  a multiplicative component comprised 
of an exponential
cutoff is combined  with our model, the E-fold energies  are in 
the expected  range of  
greater than 20  keV.  When it is applied to an additive model including a 
powerlaw, the cutoff can often extend
below 10 keV, resulting  in unreasonably low values for the inferred photon index.
As a result, the implementation of such phenomenological 
models makes it  much harder to correctly identify
the spectral state of the source, which is a central task of our
study. Additionally,   a more important property 
of the model, is that it
self-consistently calculates the normalization of the ``seed''
spectral component, which is expected to be a robust mass accretion rate
indicator. We further note that the Comptonized fraction 
is properly evaluated  by
the BMC model. 

\section{Index Saturation and Black Hole Mass Determination}

\subsection{Basic Physical Motivation}
We considered the Cyg X-3 spectra approximately resembling the galactic black-hole
low-hard state (LHS) and intermediate state (IS) 
for the basis of our study; see e.g. \cite{mr}.  We note that only a small part of
the disk emission component is seen directly. The energy spectrum is
dominated by a Comptonization component very often  approximated by a power law at energies
above $\sim 5$ keV. To
calculate the total normalization of the ``seed'' blackbody disk 
component we model the spectrum using the model described in the previous section.
As detailed in section 2, we argue that the disk
emission normalization calculated using this approach produces a more robust
correlation than that one obtains using additive models.

\subsection{Detailed Physical Interpretation}
The presence of index saturation in various black-hole -ray binaries -
Cyg X$-$1, GRO J1655$-$40, GRS 1915$+$105, GX339$-$4, H 1743$-$322, 4U
1543$-$47, and XTE J1650$-$500 $-$  has been demonstrated previously
(ST07, ST09, TS09). The self-similar evolutionary tracks of those analyses
have led to black hole (BH) mass estimates which are generally in good 
agreement with dynamical determinations.  The basic idea involves a
Compton-scattering cloud. Its scale, temperature and optical depth
regulate the nature of the emergent spectrum. We postulate that such a
cloud is a natural consequence of an adjustment from a Keplerian flow
to a turbulent, innermost sub-Keplerian boundary near the central
compact object. This barrier ultimately leads to the formation of a
transition layer between these regions.  This interpretation predicts
two phases or states dictated by the photon upscattering produced in
the transition layer: (1) an optically thin and very hot ($kT_e\sim$50
keV)  medium producing photon upscattering via thermal Comptonization;
the photon spectrum index $\Gamma\sim 1.5-1.7$ for this state is
dictated by gravitational energy release and Compton cooling in an
optically thin shock near the adjustment radius; (2) an optically
thick and relatively cool medium ($kT_e\sim 5-10 $ keV); the index for
this state, $\Gamma\gax$2.1 is determined by soft-photon upscattering
and photon trapping in converging flow into the BH. Details are presented
in ST09 and references therein.

 \subsection{Scaling Method}

The scaling technique is based on the empirical 
correlation patterns observed during spectral transitions 
which effectively form a one-parameter space; the mass of the BH 
defines a specific correlation curve (ST07, ST09). After 
scalable state transition episodes are
identified for two sources, the correlation pattern ${ \Gamma}$-$N_{bmc}$ for a reference
transition is parameterized in terms of the analytical function 
in Eq. 1. 
By fitting this functional form to the correlation pattern,
we find a set of parameters $A$, $B$, $D$,  $N_{tr}$, and $\beta$ that
represent a best$-$fit form of the function $F$ for a particular correlation curve:
 \begin{equation}
 F(x)=A-D~B\ln\{\exp[(1.0-(N/N_{tr})^\beta))/D]+1\}
 \label{f(x)_equation}
 \end{equation}

For $N\gg N_{tr}$, the correlation function $F (x)$ converges to a
constant value $A$. Thus, $A$ is the value of the index saturation
level, $\beta$ is the power-law index of the left-hand portion of the curve 
and $N_{tr}$ is the value at
which the index transitions, i.e. levels off. Parameter $D$ controls how smoothly the 
fitted function saturates to $A$. We scale the data to a template by applying
a transform $N\rightarrow s_N \times N$ until the best fit is found.
We note, that the scaling fit to determine $s_N$ has to be performed 
with all the parameters describing the shape of the pattern fixed in order to
comply with the pattern scalability requirement. 
We then use the $\Gamma-N_{bmc}$  
data points of GRS 1915$+$105, which has a reasonably
good dynamical mass determination of 
$14\pm4M_{\odot}$, \citep{gre01}, which is consistent with the value of
$15.6\pm1.5M_{\odot}$ of ST07. We adopt the latter value in our subsequent 
analysis. The reference data set is well represented 
by the function defined in equation (\ref{f(x)_equation}).
We then fit  $F(x)$ varying the parameters 
$A$,~$N_{tr}$,~$\beta$ to best represent the 
GRS 1915+105 measurements. To test for self consistency we have used both the 
least-square and chi-squared  statistical
minimization methods to determine the best fit parameter values. The
least-square method uses the values of index and normalization with equal weighting,
while in the chi-square minimization the data
are weighted by the inverse error square values. Because data points have different
error bars due to different exposure, power law fraction, etc., and because
the points can be clustered, the least-square method
is less biased. Moreover, the chi-square method results in a very 
high reduced chi-square statistics due to
large scatter in the data, which makes it less robust during the error calculation. 
We therefore select the least-squares method as our primary minimization technique.
Parameters  D and B are not well constrained  by the data.
We fix those parameters at 0.3 and 1.7 respectively based on 
our previous experience with  correlation parameterization (ST09). 
 For free parameters we obtain  
the following best-fit values with the least-square method:
$A=3.02\pm0.04$ (saturation level), $N_{tr}=0.138\pm0.004$,
$\beta=2.03\pm0.14$. 
We then obtain the scale factor $s_N=0.85\pm 0.07$ (see Fig. 4). 
We use this  value for our BH mass calculations in Cyg X-3.
We note that the scale factor obtained by the chi-square 
minimization approach,  $s_N=0.75\pm 0.08$,
leads to higher BH mass estimates.
are consistent within uncertainties. In this sense the 
least-squares method we chose represents the more conservative
of the two approaches. According to the scaling  laws (see equations  8-9 in ST09),
the expression for BH mass in Cyg X-3 is:
\begin{equation}
m_{Cyg x-3}=m_{1915}\frac{f_G}{s_N}\left(\frac{d_{Cyg X-3}}{d_{1915}}\right)^2.
\end{equation} 

Using published mass and distance determinations for GRS~1915+105
 of $d_{1915}=12.1\pm0.8$ kpc \citep{gre01}, 
$M_{1915}/M_\odot= 15.6\pm1.5$ (ST07) along with the 
value of $s_N$ determined in our least-squares analysis
leads to a BH mass in Cyg X-3 of 
$10.8\pm3.4$ M$_\odot$ and $6.2\pm2.0$ M$_\odot$ for Cyg X-3 distance of
9.3 and 7.2 kpc respectively. In the above calculation we 
used $f_G=1.0$ as discussed below. Thus over the full range
of parameters and uncertainties considered we
cannot justify the mass of the X-ray star being less than 4 solar 
masses. This seems to rule out the possibility that a neutron star is 
the compact object in Cyg X-3. Accepting the recent distance estimate 
of 9.0 kpc (\citep{pred00}) we estimate a compact object mass of $10.1\pm3.2$.

We note that the index-normalization method is subject to a several 
limitations that do not effect the index-QPO method. First, there is a 
dependence on the relative inclination angles of the binary system and
that of the template system. Second, it is not always the case that 
self-similar patterns are seen in the data. For example, while our
Cyg X-3 results were easily reconcilable with GRS 1915+105, the data 
could not be fit to an analogous template derived from XTE J1550-564. 
Finally, the relative source distances and geometry come into play, so that our
mass determination is dependent on a factor $f_G$. In the case of a
spherical coronal geometry
 $f_G$ is unity. This is a reasonable approximation for the
case of low-hard and intermediate spectral-state BHs. The ST09 results 
confirm that for most cases when the 
system inclinations are unknown the $f_G=1.0$ works very well.
In the unlikely case of a flat disk geometry $f_G$ is the ratio of inclination
angle cosines between a reference source and the target source.

\section{Discussion} 
The implications our mass estimate of $\sim10~M_{\odot}$ for the compact 
object mass is interesting in
several regards. The formation of close massive black-hole binaries is
a challenge for the theory of binary evolutionary, notably in cases
with massive donor companions. Our result suggests that Cyg X-3 not
only falls into that category, but with its remarkably short 4.8-hour period greatly
exacerbates the problem. Other notable examples to date are IC10
X-1 (1.4-day period), which is the other case of  a likely BH - WR
binary and M33 X-7 which is believed to contain one of the most
massive black holes, $\sim 16 M_{\odot}$, known in a binary system
orbiting a $\sim 70M_{\odot}-$star with a  3.5 day period. Commonly
accepted binary evolution models entail a common envelope mass
transfer episode.  This would naturally occur in massive short-period
binary systems such as Cyg X-3. This makes the mechanism
for the formation of a black hole with such a high mass very
problematic.  With the  $\sim5-$hr period, the system is so compact
that the radius of the progenitor star must have been larger than the
current separation between the stars. The BH progenitor must thus have
experienced severe mass loss via Roche-lobe overflow. This however
contradicts the clear need for a moderate mass-loss rate in order to
achieve such a high black-hole mass. Explaining both the high mass of
the black hole and the tight orbit simultaneously is difficult.
Recently, it has been suggested that there may exist
alternative stellar evolutionary paths, which avoid expansion during a
core H-burning phase, and instead become more luminous and move
blue-wards in the HR diagram. This would lead to the formation of
a He burning core but with a
chemically homogeneous outer envelope, e.g. \cite{demink09}. Such
scenarios have also been suggested in the context of gamma-ray burst
progenitors \citep{wh06}. 

Another aspect of our result is the implication that Cyg X-3 is
the first confirmed accretion-powered black hole associated with
a galactic gamma-ray source \citep{abdo09}. The other known gamma-ray 
binaries $-$ LS I 61+303, LS 5039 and
PSR 1259-63 $-$ are either known (the latter) or suspected to be powered
by wind-wind collision and/or propeller effects from a pulsar embedded
in a massive stellar wind environment. Furthermore, Cyg X-3 is
markedly different from these other systems in terms of its X-ray and
radio luminosities, its high-amplitude episodic variability and its
X-ray spectral energy distribution and spectral state changes. 
The Fermi data thus far obtained
indicate that the gamma-ray episodes are anticorrelated with the 
hard X-rays and correlated with the soft X-rays (although not every 
X-ray light curve inflection leads to detectable gamma-rays). On the other
hand, the radio emission is apparently correlated with the gamma rays; 
thus the radio turn-on or ``jet line'' of the usual "q" diagram 
characterizing black-hole X-ray binaries manifests itself
differently or may not be applicable in the case of Cyg X-3.

\section{Conclusions} 
We have studied a large set of data covering a number of spectral-state 
transitions and intensity variations in the galactic binary Cyg X-3.
We find that for about half of the 35 ~2-100-keV spectra analyzed the 
observations are well represented by our modeling approach - an absorbed
BMC model including a iron line and edge components and a high-energy 
cutoff term. We then
examined the correlation between the inferred photon indices of the
Comptonized spectral component and the mass accretion rate, a proxy for which
is the normalization term of our model. Based on those results
we have applied the scaling method employing previously determined results
for the well known Galactic binary GRS 1915+105 as a reference dataset. 
The calibration of our
method for GRS 1915+015 is supported by the dynamical mass determination of
\cite{gre01}. This analysis led us to a lower-limit mass determination 
of about $ 4.2M_\odot$ and a best-estimate value of
$\sim 10.1\pm3.2 M_\odot$ assuming a 9.0 kpc distance.
The previous success of this scaling method
for the BH  mass determination strongly supports our results
for Cyg X-3.

Another result of our study is that a basic prediction of the
theory of the converging inflow is supported by the observations of
the index-mass accretion rate correlation seen in various black-hole 
X-ray binaries. Specifically, we argue that the spectral-index vs $\dot m$
flattening, or saturation, seen is an observational
signature of the presence of a converging inflow. The implication is that
{\it this effect provides robust observational
evidence for the presence of a black hole in Cyg X-3.}

\clearpage
\newpage

\begin{deluxetable}{rrrrrrr} 
\tablecolumns{4} 
\tablewidth{0pc} 
\tablecaption{Inferred spectral parameters for mass determination } 
\tablehead{ 
\colhead{Obs ID}  & \colhead{MJD} & \colhead{Photon Index} & \colhead{Normalization} }  
\startdata 
20101-01-05-00 &  50612.46182 &  2.774 $\pm$  0.198 &  0.1817 $\pm$  0.0041 \\
20101-01-01-00 &  50604.74825 &  2.706 $\pm$  0.138 &  0.1974 $\pm$  0.0018 \\
20099-01-01-02 &  50500.75762 &  2.636 $\pm$  0.143 &  0.1483 $\pm$  0.0203 \\
20099-01-01-01 &  50500.01207 &  2.429 $\pm$  0.113 &  0.0927 $\pm$  0.0205 \\
20099-01-01-00 &  50495.01319 &  2.154 $\pm$  0.105 &  0.1032 $\pm$  0.0110 \\
20101-01-08-00 &  50632.67880 &  2.115 $\pm$  0.114 &  0.1280 $\pm$  0.0151 \\
40422-01-01-00 &  51404.43247 &  2.035 $\pm$  0.113 &  0.0966 $\pm$  0.0461 \\
10126-01-01-03 &  50324.66181 &  1.813 $\pm$  0.114 &  0.0524 $\pm$  0.0033 \\
10126-01-01-02 &  50322.25200 &  1.522 $\pm$  0.084 &  0.0696 $\pm$  0.0031 \\
10126-01-01-05 &  50325.65911 &  1.487 $\pm$  0.120 &  0.0629 $\pm$  0.0044 \\
10126-01-01-04 &  50323.65572 &  1.438 $\pm$  0.106 &  0.0673 $\pm$  0.0043 \\
20099-02-01-00 &  50717.31225 &  1.373 $\pm$  0.098 &  0.0709 $\pm$  0.0063 \\
30082-04-02-00 &  50950.94512 &  1.258 $\pm$  0.058 &  0.0622 $\pm$  0.0133 \\
10126-01-01-01 &  50321.14285 &  1.249 $\pm$  0.091 &  0.0689 $\pm$  0.0104 \\
20101-01-09-00 &  50652.62494 &  1.230 $\pm$  0.082 &  0.0936 $\pm$  0.0232 \\
\enddata 
\end{deluxetable} 

\clearpage
\newpage

\begin{figure} [ptbptbptb]
\includegraphics[scale=0.7, angle=-90]{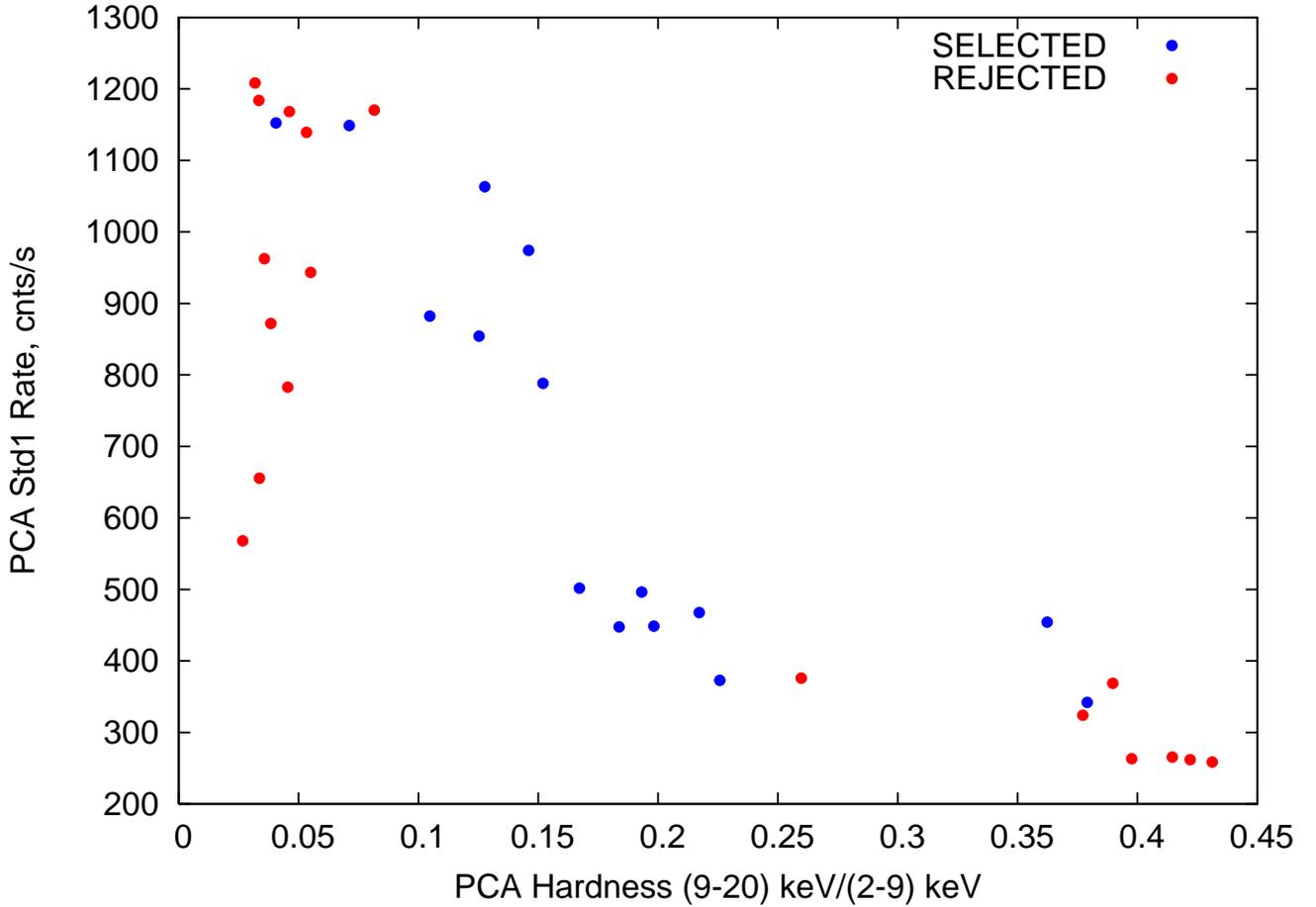}
\caption{ Hardness-intensity plot for the RXTE data considered. The data
which were used in our mass-determination procedure are plotted as the blue points, 
while the cases where our model did not fit the data satisfactorily are red. 
Note that the red points are generally at the extremes in terms of 
spectral state configuration of the binary.  }
\end{figure}

\clearpage
\newpage

\begin{figure}[ptbptbptb]
 \includegraphics[scale=0.7, angle=0]{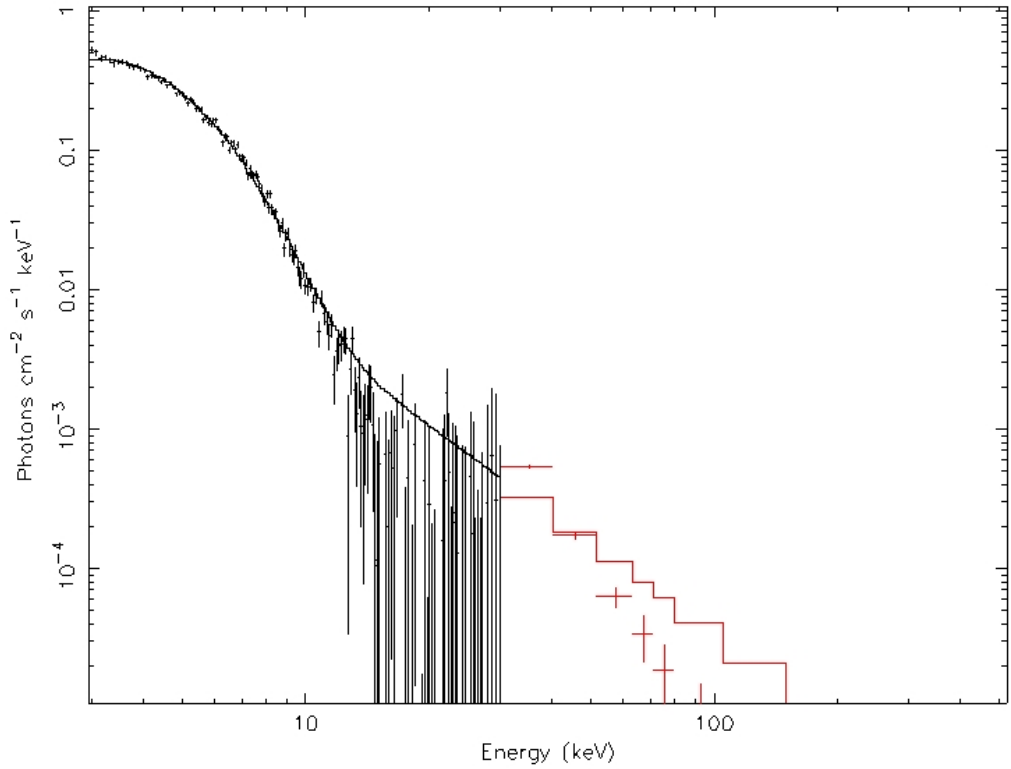}
\caption{Example of an extreme soft state case, not well suited for
our mass-estimation analysis. The hard-powerlaw index
parameter is clearly poorly constrained by the data for this 
spectral-state configuration of the source.  The corresponding 
data would lead to a point in the far left-hand side of Figure 1.
 }
\end{figure}

\clearpage
\newpage

\begin{figure}[ptbptbptb]
\includegraphics[scale=0.7, angle=0]{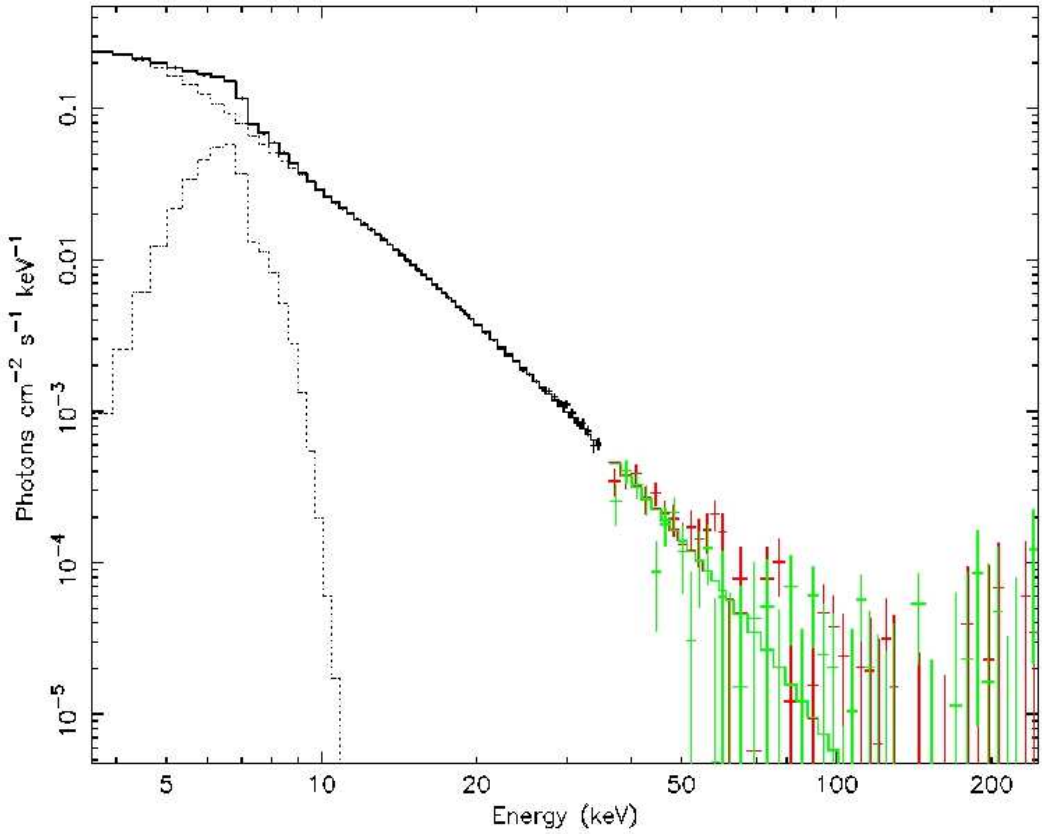}
\caption{Example of one of our satisfactory model fits to the data as described in section 2. 
Here the spectral index is hard, $\sim$1.5, well below the saturation
level but within the expected range. In this case the model fits the data at 
an acceptable level and the mass-estimation parameters are well defined.  }
\end{figure}

\clearpage
\newpage

\begin{figure}[ptbptbptb]
\includegraphics[scale=0.7, angle=-90]{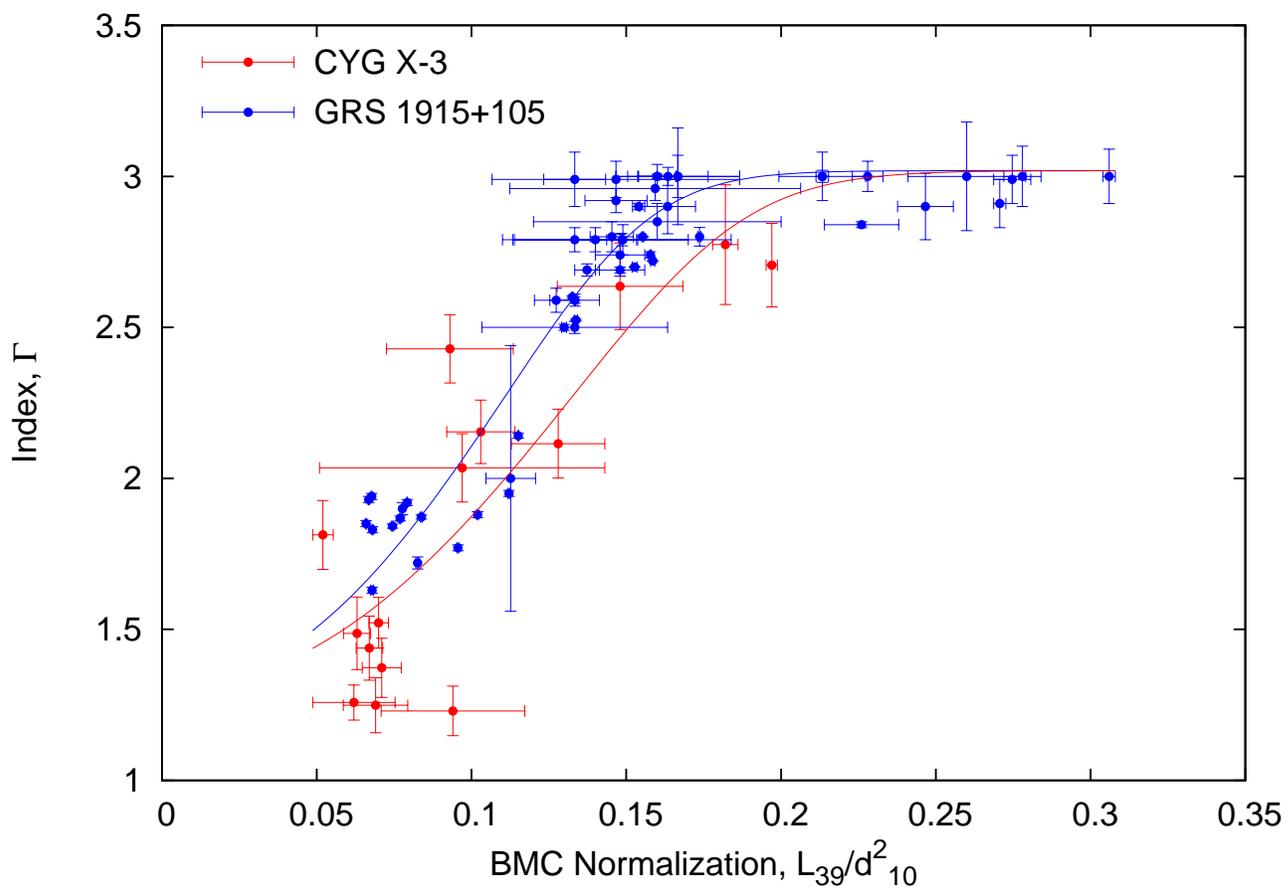}
\caption{Spectral index $\Gamma-$normalization diagram for 
Cyg X-3 (red points) in
 comparison to the our reference source, GRS 1915+105 (blue points). 
The normalization term, derived from our model, scales with 
the mass-accretion rate. The smooth curves represent the best 
fits of the functional form of equation 1 to the respective 
datasets.
The derived scaling
 parameters, as discussed in section 3.3, lead to an estimated mass
 of for the compact object in Cyg X-3 consistent with a black hole.  }
\end{figure}


\begin{thebibliography}{}

\bibitem[Abdo et al., (2009)]{abdo09}
Abdo. A. A. et al., 2009, Nature, 

\bibitem[de Mink et al. (2009)]{demink09}
de Mink,  S.E. et al. 2009, A\&A, 507, L1

\bibitem[Giacconi et al. (1967)]{gia67} 
Giacconi, R,  Gorenstein, P., Gursky, H., \& Waters, J. R. 1967, \apj, 148, L119 

\bibitem[Greiner, Cuby \& McCaughrean (2001)]{gre01} Greiner, J.,
  Cuby, J.G., \& McCaughrean, M.G. 2001, Nature, 414, 522

\bibitem[Hanson, Still \& Fender]{hsf00} Hanson, M.M., Still, M.D., \&
 Fender, R.P., 2000, ApJ, 541, 308

\bibitem[Hjalmarsdotter et al. (2009)]{hz09} Hjalmarsdotter, L., Zdziarski, A.A., 
Szostek, A., \& Hannikainen, D.C. 2009, MNRAS, 392, 251

\bibitem[Ling, Zhang \& Tang (2009)]{lzt09}
Ling,  Z., Zhang, S.N. \& Tang, S. 2009, \apj,  695, 1111

\bibitem[McClintock \& Remillard (2006)]{mr} McClintock, J. \&
  Remillard, R. 2006,  in Compact Stellar X-ray Sources,
  eds. W. H. G. Lewin \& M. van der Klis (Cambridge: Cambridge
  Univ. Press), p. 157

\bibitem[McCollough et al. (1999)]{McC99}
McCollough, M. L. et al.  (1999), \apj, 512, 951

\bibitem[Predehl et al. (2000)]{pred00} 
Predehl,  et al.  2000,  A\&A,  357, L25 

\bibitem[Remillard \& McClintock(2006)]{rm} Remillard,  R.~A., \&
  McClintock, J.~E.\ 2006, \araa, 44, 49 

\bibitem[Schmutz, W., Geballe, T.R., \& Schild, H.]{sgs96} Schmutz, W., 
 Geballe, T.R., \& Schild, H., 1996, A\&A, 311, L25

\bibitem[Shaposhnikov \& Titarchuk (2009)]{st09}  Shaposhnikov, N., \&
  Titarchuk, L. 2009, \apj, 669, 453 (ST09)

\bibitem[Shaposhnikov \& Titarchuk (2007)]{ST07}  Shaposhnikov, N., \&
  Titarchuk, L. 2007, \apj, 663, 445 (ST07)

\bibitem[Shaposhnikov \& Titarchuk (2006)]{ST06}  Shaposhnikov, N., \&
  Titarchuk, L. 2006, \apj, 643, 1098 (ST06)

 \bibitem[Shrader \& Titarchuk (1999)]{st99} Shrader, C., \&
   Titarchuk, L.G. 1999,  ApJ, 521, L121 

 \bibitem[Strohmayer \& Muchotzky (2009)]{sm09} Strohmayer, T.E., \&
   Mushotzky, R.F., 2009,  ApJ, 703, 1386

\bibitem[Titarchuk \& Seifina (2009)]{ts09} Titarchuk, L. \& Seifina,
  E.  2009, \apj,  706, 1463  

\bibitem[Titarchuk, Laurent \& Shaposhnikov  (2009)]{tls09} Titarchuk, L., Laurent, P. \&
  Shaposhnikov, N.  2009, \apj, 700, 1831 

\bibitem[Titarchuk, Lapidus \& Muslimov (1998)]{tlm98} Titarchuk, L.,
  Lapidus, I.I. \& Muslimov, A. 1998, \apj,  499, 315 (TLM98)

\bibitem[Titarchuk, Mastichiadis \& Kylafis (1997)]{bmc}  Titarchuk,
  L., Mastichiadis, A., \& Kylafis, N. D., 1997,  \apj, 487, 834

\bibitem[Titarchuk \& Shaposhnikov  (2008)]{ts08} Titarchuk, L. \&
  Shaposhnikov, N.   2008, \apj,  678, 1230

\bibitem[Titarchuk, Shaposhnikov \& Arefiev  (2007)]{tsa07} Titarchuk,
  L.,  Shaposhnikov, N. \& Arefiev, V.  2007, \apj,  660, 556

\bibitem[van Kerkwijk (1992)]{vankerk92}
van Kerkwijk,  M.H. et al. 1992, Nature, 355, 703

\bibitem[Vilhu, Hannikainen, McCollough, \& Koljonen (2009)]{vhhmk09}
 Vilhu, O., Hannikainen, D.C., McCollough, M., \& Koljonen, K., 2009, A\&A, 501, 679

\bibitem[Woosley \& Heger  (2006)]{wh06}
Woosley,  S. E. \& Heger, A.  2006 \apj, 637, 914

\bibitem[Zdziarski et al. (2009)]{zdz09}
Zdziarski, A.A., Misra, R. \& Gierlinski, M. 2009, MNRAS, in press [arXiv:0905.1086]

\end{thebibliography}
\end{document}